\documentclass{myarticle}

\usepackage{txfonts}
\usepackage[colorlinks]{hyperref}
\usepackage{tikz}

\usetikzlibrary{positioning}

\newcommand{\n}{\noindent}
\newtheorem{Theorem}{Theorem}
\newtheorem{Proposition}{Proposition}

\begin{document}

\title{F-index of Total Transformation Graphs}

\author{Nilanjan De\corref{cor1}}
\ead{de.nilanjan@rediffmail.com}

\address{Department of
Basic Sciences and Humanities (Mathematics),\\ Calcutta Institute of Engineering and Management, Kolkata, India.}
\cortext[cor1]{Corresponding Author.}

\begin{abstract}
The F-index of a graph $G$ is the sum of the cubes of the degrees of the vertices of $G$. In this paper, explicit expressions for the F-index of different transformation graphs of type ${{G}^{xyz}}$ with $x, y, z\in \{-, + \}$ are obtained. F-index for semitotal point graph and semitotal line graph are also obtained here.
\end{abstract}

\maketitle
\section{Introduction}

Chemical graph theory is a branch of mathematical chemistry  where molecular structures are modeled as molecular graphs. A molecular graph is a simple unweighted, undirected graph where the vertices correspond to atoms and edges correspond to the bonds between them. A topological index is a numeric quantity which depends on the topology of a molecular graph. More precisely, it is a real number $Top(G)$ derived from the graph $G$ such that for every graph $H$ isomorphic to $G$, $Top(G)=Top(H)$. Topological indices correlates the physico-chemical properties of molecular graphs and are used for studying quantitative structure-activity (QSAR) and structure-property (QSPR) relationships for predicting different properties of chemical compounds and biological activities. In chemistry, biochemistry and nanotechnology, different topological indices are found to be useful in isomer discrimination, QSAR, QSPR and pharmaceutical drug design.

Suppose $G$ be a simple connected graph and $V(G)$ and $E(G)$ respectively denote the vertex set and edge set of $G$. Let, $n$ and $m$ respectively denote the number of vertices and edges of $G$.  Let, for any vertex ${v}\in V(G)$, ${{d}_{G}}(v)$ denotes its degree, that is the number of neighbors of $v$ and $N(v)$ denotes the set of vertices which are the neighbors of the vertex $v$, so that $|N(v)|={{d}_{G}}(v)$.

Among various types of topological indices, degree based topological indices are most widely used indices and those have great applications in chemical graph theory. Two of the oldest and well known topological indices are the first and second Zagreb indices, which were first introduced by Gutman et al. in 1972 \cite{gutm72}, where they have examined the dependence of the total $\pi$-electron energy on molecular structure.

The first and second Zagreb indices of a graph $G$ are denoted by $M_1(G)$ and $M_2(G)$ and are respectively defined as

\[{{M}_{1}}(G)=\sum\limits_{v\in V(G)}{{{d}_{G}}{{(v)}^{2}}}=\sum\limits_{uv\in E(G)}{[{{d}_{G}}(u)+{{d}_{G}}(v)]}\] and \[{{M}_{2}}(G)=\sum\limits_{uv\in E(G)}{{{d}_{G}}(u){{d}_{G}}(v)}.\]

These  indices  are  extensively  studied  in  chemical and mathematical literature. Interested  readers  are  referred to \cite{kha09,fath11,zho07,zho05,xu15,das15} for some recent results on the topic. In the same paper, where Zagreb indices were introduced, another topological index, was also shown to influence total $\pi$-electron energy. This index was defined as sum of cubes of degrees of the vertices of the graph. However, this index never got attention except recently Furtula et al. \cite{fur15} studied this index and establish some basic properties of this index and showed that the predictive ability of this index is almost similar to that of first Zagreb index and for the entropy and acetic factor, both of them yield correlation coefficients greater than 0.95. They named this index as  ``forgotten topological index" or ``F-index". Throughout this paper, we call this index as F-index and denote it by $F(G)$. So,
\begin{eqnarray}
F(G)=\sum\limits_{v\in V(G)}{{{d}_{G}}{{(v)}^{3}}}=\sum\limits_{uv\in E(G)}{[{{d}_{G}}{{(u)}^{2}}+{{d}_{G}}{{(v)}^{2}}]}.
\end{eqnarray}
Recently, the present author have studied this index for different graph operations \cite{de16a}. They have also introduced its coindex version in \cite{de16b}. In \cite{abd15}, Abdo et al. have investigated the extremal trees with respect to the F-index.

\section{Preliminaries, Definitions and Notations}

Transformation graphs converts the information from the original graph into new transformed structure. Thus if is it possible to find out the given graph from the transformed graph, then such operation may be used to figure out structural properties of the original graph considering the transformation graphs (for details see \cite{wu01}). In different mathematical literature, several graph transformation have been considered where the vertex set of the transformed graph is equal to $V(G)\cup E(G)$. The best known type of such graphs is the total graph. The total graph $T(G)$ of a graph $G$ is  the  graph  whose  vertex  set  is $V(G)\cup E(G)$ and any two vertices of $T(G)$ are adjacent if and only if they are either incident or adjacent in $G$ \cite{yan07,nd14a,nd15,nd15a}. Two other kinds of transformation graphs were introduced by Sampathkumar in \cite{sam1,sam2,sam3} and were named as semitotal-point graph and semitotal-line graph. The semitotal-point graph of a graph $G$ is denoted by ${{T_1}}(G)$ and any two vertices $u,v\in V({{T}_{1}}(G))$ are adjacent in ${{T_1}}(G)$ if and only if either both $u$ and $v$ are adjacent vertices in $G$ or one is a vertex of $G$ and the other is an edge of $G$ incident to the former. So ${{T}_{1}}(G)$ has total $(m+n)$ vertices and $3m$ number of edges. The semitotal-line graph of a graph $G$ is denoted by ${{T_2}}(G)$ and any two vertices $u,v\in V({{T}_{2}}(G))$ are adjacent in ${{T_2}}(G)$ if and only if either both $u$ and $v$ are adjacent edges in $G$ or one is a vertex of $G$ and other is an edge of $G$ which is incident to it. So ${{T}_{1}}(G)$ has total $(m+n)$ vertices and $(\frac{1}{2}{{M}_{1}}(G)+m)$ number of edges.

Let  $x, y, z$ be three variables taking values either $+$ or $-$. Wu and Meng \cite{wu01} generalized the concept of total graphs to a total transformation graph ${{G}^{xyz}}$ with $x, y, z\in \{-, + \}$. In this paper we consider this type of total transformation graphs and hence we consider 3-permutation of the set ${+,-}$.
The transformation graphs of $G$ has the vertex set $V(G)\cup E(G)$ so that $|V({{G}^{xyz}})|=(n+m)$ and for any two vertices $u$, $v$ of the transformation graph, the associatively is $+$ if they are adjacent or incident in $G$  and  the associatively is $-$ if they are neither adjacent nor incident in $G$. Hence, for any two vertices $u,v\in V({{G}^{xyz}})$, $u$ and $v$ are adjacent in ${{G}^{xyz}}$, if and only if

(i) $u,v\in V(G)$, $u$, $v$ are adjacent in $G$ if $x = +$ and $u$, $v$ are not adjacent in $G$ if $x = -$.

(ii) $u,v\in E(G)$, $u$, $v$ are adjacent in $G$ if $y = +$ and $u$, $v$ are not adjacent in $G$ if $y= -$.

(iii) $u\in V(G)$ and $v\in E(G)$, $u$, $v$ are incident in $G$ if $z= +$ and $u$, $v$ are not incident in $G$ if $z= -$.

Since, there are eight distinct 3-permutation of ${+,-}$, so we can construct eight different transformation of a given graph $G$ of type ${{G}^{xyz}}$ where $xyz \epsilon \{+ + +, - - -, + + -, - - +, + - -, - + +, + - +, - + - \}$. From construction, it is clear that the graphs ${{G}^{+++}}$ is isomorphic to the total graph $T(G)$.
Also, for a given graph $G$, ${{G}^{+++}}$ and ${{G}^{---}}$, ${{G}^{++-}}$ and ${{G}^{--+}}$, ${{G}^{-++}}$ and ${{G}^{+--}}$, ${{G}^{+-+}}$ and ${{G}^{-+-}}$ are the pairs of complementary graphs.
The study of basic properties of these  transformation graphs were studied in \cite{xu08,yi09,wu01,zha02} where as different results of various topological indices of these transformations can be seen in \cite{hos14,bas15}. 
The topological indices, defined as the sum of contributions of all
pairs of vertices, can be expressed in terms of contributions of
edges as follows (\cite{gutm15,dos11})
\[\sum\limits_{u\in V(G)}{{{F}_{1}}(u)}=\sum\limits_{uv\in E(G)}{\left[ \frac{{{F}_{1}}(u)}{d(u)}+\frac{{{F}_{1}}(v)}{d(v)} \right]}\]
Thus if ${{F}_{1}}(u)={{d}_{G}}{{(u)}^{4}}$, then we define
\begin{eqnarray}
\sum\limits_{v\in V(G)}{{{d}_{G}}{{(v)}^{4}}}=\sum\limits_{uv\in E(G)}{[{{d}_{G}}{{(u)}^{3}}+{{d}_{G}}{{(v)}^{3}}]}={{\xi }_{4}}(G).
\end{eqnarray}
One of the redefined versions of Zagreb index is defined as
\begin{eqnarray}
{Re}Z{{G}_{3}}(G)=\sum\limits_{uv\in E(G)}{{{d}_{G}}(u){{d}_{G}}(v)[{{d}_{G}}(u)+{{d}_{G}}(v)]}.
\end{eqnarray}
For different recent study of these index see \cite{ran13,ran16,gao16,xu12}. In this paper we use this index to express the transformation graphs of a graph $G$. The following proposition was derived in \cite{de16b} and will be needed in the present study.

\begin{Proposition}
Let $G$ be a graph of order $n$ and size $m$, then
\[F(\bar{G})=2{{(n-1)}^{2}}(\bar{m}-2m)+3(n-1){{M}_{1}}(G)-F(G).\]
\end{Proposition}

Different composite and transformation graphs play a very important role  in  mathematical  chemistry,  since  some  chemically interesting graphs can be obtained from some simpler graphs by different graph operations. In this paper we derive explicit expressions for the F-index of different total transformation graphs including semitotal point graph and semitotal line graph.

\section{Main Results}

We first determine the F-index of the semitotal-point graph ${{T}_{1}}(G)$ and semitotal-line graph ${{T}_{2}}(G)$.

\begin{Theorem}
Let G be a graph of order n and size m. Then

(i) $F({{T}_{1}}(G))=8F(G)+8m.$

(ii) $F({{T}_{2}}(G))=F(G)+{{\xi }_{4}}(G)+3{Re}Z{{G}_{3}}(G).$
\end{Theorem}

\n\textit{Proof.}
(i) Since, ${{d}_{{{T}_{1}}(G)}}(u)=2{{d}_{G}}(u)$, if $u\in V({{T}_{1}}(G))\cap V(G)$and ${{d}_{{{T}_{1}}(G)}}(u)=2$, if $u\in V({{T}_{1}}(G))\cap E(G)$, we have from (1)

 \[F({{T}_{1}}(G))=\sum\limits_{u\in V(G)}{{{[2{{d}_{G}}(u)]}^{3}}}+\sum\limits_{uv\in E(G)}{{{2}^{3}}}=8F(G)+8m.\]

(ii) From definition, since ${{d}_{{{T}_{2}}(G)}}(u)=2{{d}_{G}}(u)$, if $u\in V({{T}_{2}}(G))\cap V(G)$ and ${{d}_{{{T}_{2}}(G)}}(u)=2$, if $u\in V({{T}_{2}}(G))\cap E(G)$, we get from (1)
\begin{eqnarray*}
F({{T}_{2}}(G))&=&\sum\limits_{u\in V({{G}^{+-}})}{{{d}_{{{G}^{++}}}}{{(u)}^{3}}}\\
             &=&\sum\limits_{u\in V({{G}^{++}})\cap V(G)}{{{d}_{{{G}^{++}}}}{{(u)}^{3}}}+\sum\limits_{u\in V({{G}^{++}})\cap E(G)}{{{d}_{{{G}^{++}}}}{{(u)}^{3}}}\\
             &=&\sum\limits_{u\in V(G)}{{{d}_{G}}{{(u)}^{3}}}+\sum\limits_{uv\in E(G)}{{{[{{d}_{G}}(u)+{{d}_{G}}(v)]}^{3}}}\\
             &=&F(G)+\sum\limits_{uv\in E(G)}{[{{d}_{G}}{{(u)}^{3}}+{{d}_{G}}{{(v)}^{3}}]}+3\sum\limits_{uv\in E(G)}{[{{d}_{G}}(u)+{{d}_{G}}(v)]}{{d}_{G}}(u){{d}_{G}}(v)\\
             &=&F(G)+{{\xi }_{4}}(G)+3{Re}Z{{G}_{3}}(G).
\end{eqnarray*}
Hence the result. \qed


Now, we consider F-index of the total transformation graphs of type ${{G}^{xyz}}$,  where  $x=y=z=+$.

\begin{Theorem}
Let $G$ be a graph of order $n$ and size $m$. Then
\[F({{G}^{+++}})=8F(G)+{{\xi }_{4}}(G)+3{Re}Z{{G}_{3}}(G).\]
\end{Theorem}

\n\textit{Proof.}
From the construction of ${{G}^{+++}}$ we have, for  $u\in V({{G}^{+++}})\cap V(G)$,  ${{d}_{{{G}^{+++}}}}(u)=2{{d}_{G}}(u)$  and  for $u\in V({{G}^{+++}})\cap E(G)$,  ${{d}_{{{G}^{+++}}}}(u)={{d}_{G}}(u)+{{d}_{G}}(v)$.
\begin{eqnarray*}
F({{G}^{+++}})&=&\sum\limits_{v\in V({{G}^{+++}})}{{{d}_{{{G}^{+++}}}}{{(v)}^{3}}}\\
              &=&\sum\limits_{u\in V({{G}^{+++}})\cap V(G)}{{{d}_{{{G}^{+++}}}}{{(v)}^{3}}}+\sum\limits_{u\in V({{G}^{+++}})\cap E(G)}{{{d}_{{{G}^{+++}}}}{{(v)}^{3}}}\\
              &=&\sum\limits_{v\in V(G)}{8{{d}_{G}}{{(u)}^{3}}}+\sum\limits_{uv\in E(G)}{{{[{{d}_{G}}(u)+{{d}_{G}}(v)]}^{3}}}\\
              &=&8F(G)+\sum\limits_{uv\in E(G)}{[{{d}_{G}}{{(u)}^{3}}+{{d}_{G}}{{(v)}^{3}}]}+3\sum\limits_{uv\in E(G)}{{{d}_{G}}(u){{d}_{G}}(v)[{{d}_{G}}(u)+{{d}_{G}}(v)]}
\end{eqnarray*}
from where the desired result follows.  \qed

Next, we calculate F-index of the complement of total transformation graphs of above type, that is of ${{G}^{xyz}}$,  where  $x=y=z=-$.

\begin{Theorem}
Let $G$ be a graph of order $n$ and size $m$. Then
\begin{eqnarray*}
F({{G}^{---}})&=&(3m+3n-11)F(G)-3(m+n-1)(m+n-5){{M}_{1}}(G)+6(m+n-1){{M}_{2}}(G)\\
              &&-{{\xi }_{4}}(G)-3{Re}Z{{G}_{3}}(G)+(m+n){{(m+n-1)}^{3}}-12m{{(m+n-1)}^{2}}.\\
\end{eqnarray*}

\end{Theorem}

\n\textit{Proof.}
Since, ${{G}^{---}}\cong \overline{{{G}^{+++}}}$, where the notation $\cong$ indicates the two graphs are isomorphic, the desired result follows from the Proposition 1 by applying Theorem 3 and the following results.
\[|E({{G}^{+++}})|=\frac{1}{2}[4m+{{M}_{1}}(G)]\]
\[{{M}_{1}}({{G}^{+++}})=4{{M}_{1}}(G)+2{{M}_{2}}(G)+F(G)\]
This result can also derived from direct calculation as for  $u\in V({{G}^{---}})\cap V(G)$, ${{d}_{{{G}^{---}}}}(u)=m+n-1-2{d_G}(u)$  and  for $u\in V({{G}^{---}})\cap E(G)$, ${{d}_{{{G}^{---}}}}(u)=m+n-1-({{d}_{G}}(u)+{{d}_{G}}(v))$.
\qed

In the following, we consider F-index of the total transformation graphs of type ${{G}^{xyz}}$,  where  $x=y=+$ and $z=-$.

\begin{Theorem}
Let $G$ be a graph of order $n$ and size $m$. Then
\begin{eqnarray*}
F({{G}^{++-}})&=&3(n-4)F(G)+3{{(n-4)}^{2}}{{M}_{1}}(G)+6(n-4){{M}_{2}}(G)+{{\xi }_{4}}(G)\\
              &&+3{Re}Z{{G}_{3}}(G)+n{{m}^{3}}+m{{(n-4)}^{3}}.
\end{eqnarray*}
\end{Theorem}

\n\textit{Proof.}
We have, for  $u\in V({{G}^{++-}})\cap V(G)$, ${{d}_{{{G}^{++-}}}}(u)=m$  and  for $u\in V({{G}^{++-}})\cap E(G)$, ${{d}_{{{G}^{++-}}}}(u)={{d}_{G}}(u)+{{d}_{G}}(v)+n-4$.  Thus from (1) we have

\begin{eqnarray*}
 F({{G}^{++-}})&=&\sum\limits_{v\in V({{G}^{++-}})}{{{d}_{{{G}^{++-}}}}{{(v)}^{3}}}\\
               &=&\sum\limits_{u\in V({{G}^{++-}})\cap V(G)}{{{d}_{{{G}^{++-}}}}{{(v)}^{3}}}+\sum\limits_{u\in V({{G}^{++-}})\cap E(G)}{{{d}_{{{G}^{++-}}}}{{(v)}^{3}}}\\
               &=&\sum\limits_{v\in V(G)}{{{m}^{3}}}+\sum\limits_{uv\in E(G)}{{{[{{d}_{G}}(u)+{{d}_{G}}(v)+(n-4)]}^{3}}}\\
               &=&n{{m}^{3}}+\sum\limits_{uv\in E(G)}{{{[{{d}_{G}}(u)+{{d}_{G}}(v)]}^{3}}}+3{{(n-4)}^{3}}\sum\limits_{uv\in E(G)}{[{{d}_{G}}(u)+{{d}_{G}}(v)]}\\
               &&+3(n-4)\sum\limits_{uv\in E(G)}{{{[{{d}_{G}}(u)+{{d}_{G}}(v)]}^{2}}+}\sum\limits_{uv\in E(G)}{{{(n-4)}^{3}}}\\
               &=&n{{m}^{3}}+m{{(n-4)}^{3}}+{{\xi }_{4}}(G)+3{Re}Z{{G}_{3}}(G)+3{{(n-4)}^{2}}{{M}_{1}}(G)+3(n-4)[F(G)+2{{M}_{2}}(G)].
 \end{eqnarray*}
from where the desired result follows.    \qed


Now, similarly we calculate F-index of the complement of total transformation graphs of previous type, that is of ${{G}^{xyz}}$,  where  $x=y=-$ and $z=+$.

\begin{Theorem}
Let $G$ be a graph of order $n$ and size $m$. Then
\begin{eqnarray*}
F({{G}^{--+}})&=&3(m+3)F(G)-3{{(m+3)}^{2}}{{M}_{1}}(G)+6(m+3){{M}_{2}}(G)-{{\xi }_{4}}(G)-3{Re}Z{{G}_{3}}(G)\\
              &&+n{{(n-1)}^{3}}+m{{(m+3)}^{3}}.
\end{eqnarray*}
\end{Theorem}

\n\textit{Proof.}
Again since, ${{G}^{--+}}=\overline{{{G}^{++-}}}$, the desired result follows from the Proposition 1 by applying Theorem 5 and the following results.
\[|E({{G}^{++-}})|=\frac{1}{2}[2m(n-2)+{{M}_{1}}(G)]\]
\[{{M}_{1}}({{G}^{++-}})=mn(m+n-8)+16m+2(n-4){{M}_{1}}(G)+2{{M}_{2}}(G)+F(G)\]
It can be easily seen that, this result can also derived from directly as for  $u\in V({{G}^{--+}})\cap V(G)$, ${{d}_{{{G}^{--+}}}}(u)=n-1$  and  for $u\in V({{G}^{--+}})\cap E(G)$, ${{d}_{{{G}^{--+}}}}(u)=(m+3)-({{d}_{G}}(u)+{{d}_{G}}(v))$.  \qed

Next, we consider F-index of the total transformation graphs of type ${{G}^{xyz}}$,  where  $x=-$ and $y=z=+$.

\begin{Theorem}
Let $G$ be a graph of order $n$ and size $m$. Then
\[F({{G}^{-++}})=n{{(n-1)}^{3}}+{{\xi }_{4}}(G)+3{Re}Z{{G}_{3}}(G).\]
\end{Theorem}

\n\textit{Proof.}
From definition we have, for  $u\in V({{G}^{-++}})\cap V(G)$, ${{d}_{{{G}^{-++}}}}(u)=(n-1)$  and  for $u\in V({{G}^{-++}})\cap E(G)$, ${{d}_{{{G}^{-++}}}}(u)={{d}_{G}}(u)+{{d}_{G}}(v)$.
\begin{eqnarray*}
F({{G}^{-++}})&=&\sum\limits_{v\in V({{G}^{-++}})}{{{d}_{{{G}^{-++}}}}{{(v)}^{3}}}\\
              &=&\sum\limits_{u\in V({{G}^{-++}})\cap V(G)}{{{d}_{{{G}^{-++}}}}{{(v)}^{3}}}+\sum\limits_{u\in V({{G}^{-++}})\cap E(G)}{{{d}_{{{G}^{-++}}}}{{(v)}^{3}}}\\
              &=&\sum\limits_{v\in V(G)}{{{(n-1)}^{3}}}+\sum\limits_{uv\in E(G)}{{{[{{d}_{G}}(u)+{{d}_{G}}(v)]}^{3}}}\\
              &=&n{{(n-1)}^{3}}+\sum\limits_{uv\in E(G)}{[{{d}_{G}}{{(u)}^{3}}+{{d}_{G}}{{(v)}^{3}}]}+3\sum\limits_{uv\in E(G)}{{{d}_{G}}(u){{d}_{G}}(v)[{{d}_{G}}(u)+{{d}_{G}}(v)]}
\end{eqnarray*}
from where the desired result follows.   \qed


Similarly, in the following we calculate F-index of the complement of total transformation graphs of previous type, that is of ${{G}^{xyz}}$, where  $x=+$ and $y=z=-$.

\begin{Theorem}
Let $G$ be a graph of order $n$ and size $m$. Then
\begin{eqnarray*}
F({{G}^{+--}})&=&3(m+n-1)F(G)-{{\xi }_{4}}(G)-3{Re}Z{{G}_{3}}(G)-3{{(m+n-1)}^{2}}{{M}_{1}}(G)\\
              &&+6(m+n-1){{M}_{2}}(G)+m{{(m+n-1)}^{3}}+n{{m}^{3}}.
\end{eqnarray*}
\end{Theorem}

\n\textit{Proof.}
Again since, ${{G}^{+--}}=\overline{{{G}^{-++}}}$, the desired result follows from the Proposition 1 by applying Theorem 7 and the following results.
\[|E({{G}^{-++}})|=\frac{1}{2}[n(n-1)+{{M}_{1}}(G)]\]
\[{{M}_{1}}({{G}^{-++}})=n{{(n-1)}^{2}}+2{{M}_{2}}(G)+F(G)\]
This result can also derived from direct calculation as for  $u\in V({{G}^{+--}})\cap V(G)$, ${{d}_{{{G}^{+--}}}}(u)=m$  and  for $u\in V({{G}^{+--}})\cap E(G)$, ${{d}_{{{G}^{+--}}}}(u)=(m+n-1)-({{d}_{G}}(u)+{{d}_{G}}(v))$.    \qed

Next, we consider F-index of the total transformation graphs of type ${{G}^{xyz}}$, where $x=z=+$ and $y=-$.

\begin{Theorem}
Let $G$ be a graph of order $n$ and size $m$. Then
\begin{eqnarray*}
F({{G}^{+-+}})&=&(3m+17)F(G)-3{{(m+3)}^{2}}{{M}_{1}}(G)+6(m+3){{M}_{2}}(G)-{{\xi }_{4}}(G)\\
              &&-3e{Re}Z{{G}_{3}}(G)+m{{(m+3)}^{3}}.
\end{eqnarray*}
\end{Theorem}

\n\textit{Proof.}
From definition we have, for  $u\in V({{G}^{+-+}})\cap V(G)$, ${{d}_{{{G}^{+-+}}}}(u)=2{{d}_{G}}(v)$  and  for $u\in V({{G}^{+-+}})\cap E(G)$${{d}_{{{G}^{+-+}}}}(u)=m+3-({{d}_{G}}(u)+{{d}_{G}}(v))$.

\begin{eqnarray*}
F({{G}^{+-+}})&=&\sum\limits_{v\in V({{G}^{+-+}})}{{{d}_{{{G}^{+-+}}}}{{(v)}^{3}}}\\
              &=&\sum\limits_{u\in V({{G}^{+-+}})\cap V(G)}{{{d}_{{{G}^{+-+}}}}{{(v)}^{3}}}+\sum\limits_{u\in V({{G}^{+-+}})\cap E(G)}{{{d}_{{{G}^{+-+}}}}{{(v)}^{3}}}\\
              &=&\sum\limits_{v\in V(G)}{8{{d}_{G}}{{(v)}^{3}}}+\sum\limits_{uv\in E(G)}{{{[(m+3)-({{d}_{G}}(u)+{{d}_{G}}(v))]}^{3}}}\\
              &=&8F(G)+m{{(m+3)}^{3}}-3{{(m+3)}^{2}}\sum\limits_{uv\in E(G)}{[{{d}_{G}}(u)+{{d}_{G}}(v)]}\\
                &&+3(m+3){{\sum\limits_{uv\in E(G)}{[{{d}_{G}}(u)+{{d}_{G}}(v)]}}^{2}}-{{\sum\limits_{uv\in E(G)}{[{{d}_{G}}(u)+{{d}_{G}}(v)]}}^{3}}\\
               &=&8F(G)+m{{(m+3)}^{3}}-3{{(m+3)}^{2}}{{M}_{1}}(G)+3(m+3)[F(G)+2{{M}_{2}}(G)]\\
               &&-{{\xi }_{4}}(G)-3{Re}Z{{G}_{3}}(G)
\end{eqnarray*}
from where the desired result follows.  \qed


Similarly, in the following we calculate F-index of the complement of total transformation graphs of previous type, that is of ${{G}^{xyz}}$, where $x=z=-$ and $y=+$.

\begin{Theorem}
Let $G$ be a graph of order $n$ and size $m$. Then

\begin{eqnarray*}
F({{G}^{-+-}})&=&(3n-20)F(G)+(3{{n}^{2}}-12n+12m+36){{M}_{1}}(G)+6(n-4){{M}_{2}}(G)+{{\xi }_{4}}(G)\\
              &&+3{Re}Z{{G}_{3}}(G)+{{(m+n-1)}^{2}}\{n(m+n-1)-12m\}+m{{(n-4)}^{3}}.
\end{eqnarray*}

\end{Theorem}

\n\textit{Proof.}
Again since, ${{G}^{-+-}}=\overline{{{G}^{+-+}}}$, the desired result follows from the Proposition 1 by applying Theorem 9 and the following results.

\[|E({{G}^{+-+}})|=\frac{1}{2}[m(m+7)-{{M}_{1}}(G)]\]

\[{{M}_{1}}({{G}^{+-+}})=m{{(m+3)}^{2}}-2(m+1){{M}_{1}}(G)+2{{M}_{2}}(G)+F(G)\]

This result can also derived from direct calculation as for  $u\in V({{G}^{-+-}})\cap V(G)$, ${{d}_{{{G}^{-+-}}}}(u)=m+n-1-2{d_G}(u)$  and  for $u\in V({{G}^{-+-}})\cap E(G)$, ${{d}_{{{G}^{-+-}}}}(u)=(n-4)+({{d}_{G}}(u)+{{d}_{G}}(v))$.     \qed

\section{Conclusion}

In this paper, we determine the F-index of transformation graphs in terms of different graph invariants and showed that how it changes under different transformations. The results obtained in this study illustrate the prospects of application of F-index for chemical, biological, pharmaceutical sciences.For further studies different other topological indices  of some new derived graphs can be obtained to understand the underlying topology.




\begin{thebibliography}{999}

\bibitem{kha09}
Khalifeha, M.H.; Yousefi-Azaria, H.; Ashrafi, A.R. The first and second Zagreb indices of some graph operations. {\em Disc. Appl. Math.} {bf\ 2009}, {em\ 157(4)}, 804-811.

\bibitem{gutm72}
Gutman, I.; Trinajsti\'{c}, N. {Graph theory and molecular orbitals. Total $\pi$-electron energy of alternant hydrocarbons}. {\em Chem. Phys. Lett.} {\bf 1972}, {\em 17}, 535-538.

\bibitem{fur15}
Furtula, B.; Gutman, I. {A forgotten topological index.} {\em J. Math. Chem.} {\bf 2015}, {\em 53(4)}, 1184-1190.


\bibitem{de16a}
De, N.; Nayeem, S.M.A.; Pal, A. {F-index of some graph operations.} {\em Discrete Math. Algorithm. Appl.} {\bf 2016}, doi :10.1142/S1793830916500257.

\bibitem{abd15}
Abdoa, H.; Dimitrov D.; Gutman, I. {On extremal trees with respect to the F-index.} {\em arXiv:1509.03574} {\bf 2015}.


\bibitem{yan07}
Yan, W.; Yang, B.Y.; Yeh, Y.N. {The behavior of Wiener indices and polynomials of graphs under fivegraph decorations.} {\em Appl. Math. Lett.} {\bf 2007}, {\em 20}, 290-295.


\bibitem{fath11}Fath-Tabar, G.H. Old and new Zagreb indices of graphs. {\em MATCH Commun. Math. Comput. Chem.} {\bf 2011}, {\em 65},  79-84.





\bibitem{zho07}
Zhou, B. Upper bounds for the Zagreb indices and the spectral radius of series-parallel graphs. {\em Int. J. Quantum Chem.} {\bf 2007}, {\em 107}, 875-878.


\bibitem{zho05}Zhou, B.; Gutman, I. Further properties of Zagreb indices, {\em MATCH Commun. Math. Comput. Chem.}, {\bf 2005} {em\ 54} , 233-239.

\bibitem{xu15}Xu, K.; Tang, K.; Liu, H.; Wang, J.  The Zagreb indices of bipartite graphs with more edges. {\em J. Appl. Math. \& Informatics} {\bf 2015}, {\em 33}, 365-377.

\bibitem{das15}Das, K.C.; Xu, K.; Nam, J. On Zagreb indices of graphs, {\em Front. Math. China} {\bf 2015}, {\em 10}, 567-582.

\bibitem{ran13}Ranjini, P.S.; Lokesha, V.; Usha, A. Relation between phenylene and hexagonal squeeze using harmonic index. {\em Int. J. of Graph Theory} {\bf 2013}, {\em 1}, 116-121.

\bibitem{gutm15}Gutman I. Edge-decomposition of topological indices. {\em Iranian Journal of Mathematical Chemistry} {\bf 2015} {\em 6(2)}, 103-108.

\bibitem{dos11}Dosli$\acute{\rm c }$, T.; Reti, T.; Vukicevic, D. On the vertex degree indices of connected graphs. {\em Chem. Phys. Lett.} {\bf 2011} {\em 512}, 283-286.

\bibitem{hos14}Hosamani, S.M.; Gutman, I. Zagreb indices of transformation graphs and total transformation graphs. {\em Appl. Math. Comput.} {\bf 2014}, {\em 247}, 1156-1160.


\bibitem{bas15}Basavanagoud, B.; Gutman, I.; Desai, V.R. Zagreb indices of generalized transformation graphs and their complement. {\em Kragujevac J. Sci.} {\bf 2015}, {\em 37}, 99-112.


\bibitem{sam1}Sampathkumar, E.; Chikkodimath, S.B. Semitotal graphs of a graph - I. {\em J. Karnatak Univ. Sci.} {\bf 1973}, {\em 18}, 274-280.

\bibitem{sam2}Sampathkumar, E.; Chikkodimath, S.B. Semitotal graphs of a graph - II. {\em J. Karnatak Univ. Sci.} {\bf 1973}, {\em 18}, 281-284.

\bibitem{sam3}Sampathkumar, E.; Chikkodimath, S.B. Semitotal graphs of a graph - III. {\em J. Karnatak Univ. Sci.} {\bf 1973}, {\em 18}, 285-296.

\bibitem{xu08}Xu, L.; Wu, B. Transformation graph ${G^{-+-}}$. {\em Discrete Math.} {\bf 2008}, {\em 308}, 5144-5148.

\bibitem{yi09}L. Yi., B. Wu, The transformation graph ${G^{++-}}$, {\em Aust. J. Comb.} {\bf 2009}, {\em 44}, 37-42.

\bibitem{wu01}Wu, B.; Meng, J. Basic properties of total transformation graphs. {\em J. Math. Study} {\bf 2001}, {\em 34}, 109-116.

\bibitem{zha02}Zhang, Z.; Huang, X. Connectivity of transformation graph ${G^{+-+}}$. {\em Graph Theory Notes of New York} {\bf 2002}, {\em XLIII}, 35-38.

\bibitem{xu12}Xu, X. Relationships between harmonic index and other topological indices. {\em Appl. Math. Sc.} {\bf 2012}, {\em 6(41)}, 2013-2018.

\bibitem{ran16}Ranjini, P.S.; Usha, A.; Lokesha, V.; Deepika, T. Harmonic index, redefined Zagreb indices of dragon graph with complete graph. {\em Asian J. Math. Comp. Res.} {\bf 2016}, {\em 9(2)}, 161-166.

\bibitem{gao16}W. Gao,  W. Wang  and M.R. Farahani, Topological Indices Study of Molecular Structure in Anticancer Drugs,  {\em J. Chem.} , {\bf 2016}, http://dx.doi.org/10.1155/2016/3216327.


\bibitem{de16b}
De, N.; Nayeem, S.M.A.; Pal, A. The F-coindex of some graph operations. {\em SpringerPlus} {\bf 2016}, {\em 5:221}, doi: 10.1186/s40064-016-1864-7.

\bibitem{nd14a}
De, N.; Nayeem, S.M.A.; Pal, A. Total eccentricity index of the generalized hierarchical product of graphs. {\em Int. J. Appl. Comput. Math.} {\bf 2014}, {doi:10.1007/s40819-014-0016-4}.

\bibitem{nd15a}
De, N.; Pal, A.; Nayeem, S.M.A. The irregularity of some composite graphs. {\em Int. J. Appl. Comput. Math.} {\bf 2015}, {doi:10.1007/s40819-015-0069-z}.

\bibitem{nd15}
De, N.; Pal, A.; Nayeem, S.M.A. Total eccentricity index of some composite graphs. {\em Malaya J. Mat.} {\bf 2015}, {\em 3(4)}, 523-529.

\end{thebibliography}
\end{document}